\documentstyle[12pt,epsf]{ioplppt}
\newcommand{\be}{\begin{equation}}
\newcommand{\ee}{\end{equation}} %\indent}
\newcommand{\eei}{\end{equation}\indent\indent}
\newcommand{\bc}{\begin{center}}
\newcommand{\ec}{\end{center}}
\newcommand{\ber}{\begin{eqnarray}}
\newcommand{\ear}{\end{eqnarray}}
\newcommand{\ba}{\begin{array}}
\newcommand{\ea}{\end{array}}
\newcommand{\n}{{}^{(3)}\nabla} 

\newcommand{\hs}{\,-\,}
\newcommand{\p}{\partial}

\newcommand{\dd}{{\cal D}}
\newcommand{\Hs}{{\cal H}}

\def\case#1/#2{\textstyle\frac{#1}{#2} }
\begin{document}
%%%%%%%%%%%%%%%%%%%%%%%%%%%%%%%%%%%%%%%%%%%%%%%%
%\addtolength{\textwidth}{0.2\textwidth}
%\hoffset=-37pt
\title{A Fully Covariant Description of CMB Anisotropies}
\author{Peter K. S. Dunsby}
\address{Department of Applied Mathematics, University of Cape Town, 
Rondebosch 7700, Cape Town, South Africa and} 
\address{Department of Mathematics, Statistics, and Computing Science,
Dalhousie University, Halifax, Nova Scotia, Canada}
%\date{\today}
%%%%%%%%%%%%%%%%%%%%%%%%%%%%%%%%%%%%%%%%%%%%%%%%
\begin{abstract}
%%%%%%%%%%%%%%%%%%%%%%%%%%%%%%%%%%%%%%%%%%%%%%%%
Starting from the exact non\hs linear description of matter and 
radiation, a fully covariant and gauge\hs invariant formula for 
the observed temperature anisotropy of the cosmic microwave 
background (CBR) radiation, expressed in terms of the electric ($E_{ab}$) 
and magnetic ($H_{ab}$) parts of the Weyl tensor, is obtained by 
integrating photon geodesics from last scattering to the point of 
observation today. This improves and extends earlier work by 
Russ {\it et al} where a similar formula was obtained by taking 
first order variations of the redshift. In the case of scalar (density) 
perturbations, $E_{ab}$ is related to the harmonic components of the 
gravitational potential $\Phi_k$ and the usual dominant 
Sachs\hs Wolfe contribution $\delta T_R/\bar{T}_R\sim\Phi_k$ to the 
temperature anisotropy is recovered, together with contributions 
due to the time variation of the potential (Rees\hs Sciama effect), 
entropy and velocity perturbations at last scattering and a pressure 
suppression term important in low density universes. We also explicitly 
demonstrate the validity of assuming that the perturbations are adiabatic 
at decoupling and show that if the surface of last scattering is 
correctly placed and the background universe model is taken to be a 
flat dust dominated Friedmann\hs Robertson\hs Walker model (FRW), 
then the large scale temperature anisotropy can be interpreted as being 
due to the motion of the matter relative to the surface of constant 
temperature which defines the surface of last scattering on those scales.
\end{abstract}
%%%%%%%%%%%%%%%%%%%%%%%%%%%%%%%%%%%%%%%%%%%%%%%%
%\bc PACS number(s): 98.80.Hw \ec
%%%%%%%%%%%%%%%%%%%%%%%%%%%%%%%%%%%%%%%%%%%%%%%
\section{Introduction} \label{sec:intro}   
%%%%%%%%%%%%%%%%%%%%%%%%%%%%%%%%%%%%%%%%%%%%%%%
The study of the cosmic microwave background (CMB) radiation is
the corner stone of modern Big Bang cosmology and has led to its widespread
acceptance. Over the last few years improved measurements of the temperature
spectrum and anisotropy has led to a better understanding of the origin and
evolution of large scale structure in the universe. 

The physical basis for using anisotropies in the CMB to constrain competing 
theories of galaxy formation is the Sachs\hs Wolfe effect \cite{bi:sw}. 
The basic assumption is that the observed CMB photons travel to us without 
significant interaction with matter from a redshift of about $z=1200$. At 
redshifts greater than this, the universe is ionized and photons are 
coupled to the electron\hs baryon plasma through Thompson scattering. 

The process of decoupling i.e. the transition of the CMB from a collisional 
regime to being free photons does not take place instantaneously. 
The thickness of the decoupling shell $\Delta z$ is approximately $1/15$ 
of the mean redshift, which from our point of view as observers is 
relatively narrow \cite{bi:hogan}, so for many purposes we can  
treat this shell as a sharp surface, called the {\it surface of last 
scattering} (SLS). The correct way of placing this surface is by
determining where the optical depth due to Thompson scattering is 
unity \cite{bi:panek,bi:SXEK,bi:SES,bi:adiab1}. This occurs, to first order, 
where the radiation temperature (which is equal to the matter 
temperature in the strongly coupled region prior to decoupling) 
reaches the matter ionization temperature, so if we take decoupling as 
happening essentially instantaneously, the SLS is, to good approximation, 
a surface of constant radiation temperature.

Causally connected regions at the SLS, as viewed by an observer today, 
subtend an angle $\theta\sim2\sqrt{\Omega_0}$, where $\Omega_0$ is the 
present value of the density parameter, so large scale anisotropies on 
angular scales greater than $7$ degrees are unaffected by the small scale 
physics of decoupling, and so represent primordial perturbations. These 
anisotropies arise because photons traveling from the SLS are red\hs shifted 
slightly more than they would be in a perfectly homogeneous universe as 
a result of having to climb from an increased gravitational potential due to 
density perturbations over the surface.

The calculation of CMB anisotropies on angular scales larger than a degree,
corresponding to scales larger than the Hubble radius at the time of 
decoupling, is very simple in principle, however in practice the formulation 
of a completely self\hs consistent theoretical picture has been plagued by a
misunderstanding of the observational meaning of certain temperature
perturbation measures and whether or not these measures are gauge\hs
invariant. Furthermore many approximations are made in deriving the various
contributions to the anisotropy, without adequate consideration of 
whether or not these approximations are justifiable or consistent with 
each other.

In this paper we attempt to address some of the above problems, by 
deriving from first principles a formula for the CMB anisotropy which is 
both simpler and easier to interpret than many of the usual treatments. The 
approach we take is to integrate photon geodesics from the time of last 
scattering to today, obtaining a general (model independent) formula for the 
observed temperature in a given direction in the sky and expressing the result 
in terms of covariantly defined gauge\hs invariant quantities. The 
temperature anisotropy is then given by subtracting this result for two 
independent directions corresponding to different points of emission on 
the SLS. This improves and extends earlier work by Russ {\it et al} 
\cite{bi:russetal} by (i) starting from a general non\hs linear treatment
of the geodesic equation and linearizing to obtain the {\it almost} FRW
result, rather than taking first order variations of the redshift; 
(ii) correctly defining the temperature perturbation measure 
$\delta T/\bar{T}$; and (iii) performing a two\hs fluid analysis instead
of treating the radiation as a test field on a single\hs fluid background.

Following this, we demonstrate that given a number of ``standard'' 
assumptions, the classical Sachs\hs Wolfe result is recovered and we also 
consider in detail the validity of the assumption that the perturbations 
in the total energy density of the photon\hs baryon fluid at decoupling 
are adiabatic.

Conventions on signature, Riemann and Ricci tensors are as in
\cite{bi:ellis2}, and the speed of light is taken to be unity ($c=1$).
Standard General Relativity is assumed, with Einstein's equations
in the form $G_{ab}=\kappa T_{ab}$ where $G_{ab}$ is the usual Einstein
tensor, $\kappa=8\pi G$ is the gravitational constant and $T_{ab}$ is the
energy momentum tensor of the matter. Most of the notation is the
same as in \cite{bi:BDE,bi:DBE} and any changes are stated in the text.
%%%%%%%%%%%%%%%%%%%%%%%%%%%%%%%%%%%%%%%%%%%%%%%
\section{Basic equations and notation} \label{sec:basic}
%%%%%%%%%%%%%%%%%%%%%%%%%%%%%%%%%%%%%%%%%%%%%%%
For the sake of self\hs consistency and clarity, we will first summarize the 
covariant approach to cosmology.
%%%%%%%%%%%%%%%%%%%%%%%%%%%%%%%%%%%%%%%%%%%%%%%
\subsection{The covariant approach}
%%%%%%%%%%%%%%%%%%%%%%%%%%%%%%%%%%%%%%%%%%%%%%%
As in Hawking \cite{bi:hawking} and Ellis \cite{bi:ellis1,bi:ellis2}, 
the hydrodynamic fluid 4\hs velocity (tangent to the worldlines of 
fundamental observers in the universe) is $u^a=dx^a/dt$ ($u^au_a=-1$), 
where $t$ is the proper time along the flow lines. The projection tensor
into the tangent three\hs spaces orthogonal to $u^a$ (the local rest 
frame of these observers) is 
\be
h_{ab}=g_{ab}+u_au_b\;.
\ee
The first covariant derivative of $u_a$ can be uniquely decomposed 
into four parts:
\be
u_{a;b}=\n_bu_a-\dot{u}_au_b\;,
\label{eq:uab}
\ee
where 
\be
\n_bu_a=\sigma_{ab}+\omega_{ab}+\case{1}/{3}\Theta h_{ab}\;,
\label{eq:cov}
\ee
and $\n_a$ is the spatial gradient operator, orthogonal to $u^a$. Here 
$\Theta=u^a{}_{;a}$ is the volume expansion, $\sigma_{ab}=\sigma_{(ab)}$ 
is the shear tensor ($\sigma_{ab}u^a=\sigma^a{}_a=0$),
$\omega_{ab}=\omega_{[ab]}$ is the vorticity ($\omega_{ab}u^b=0$) and 
$\dot{u}^a=u_{a;b}u^b$ is the acceleration (the dot denotes a proper time 
derivative). It is useful to introduce a {\it length scale} along 
the fluid flow lines by the relation
\be
\frac{\dot{a}}{a}=\frac{1}{3}\Theta=H\;.
\ee
When the universe is an exact FRW spacetime, $H$ is just the usual 
Hubble parameter. In general, however, the evolution equation for the 
expansion $\Theta$ is the Raychaudhuri equation
\be
\dot{\Theta}+\case{1}/{3}\Theta^2+2(\sigma^2-\omega^2)
-\n^a\dot{u}_a-\dot{u}_a\dot{u}^a+\case{1}/{2}\kappa (\mu+3p)=0\;,
\label{eq:theta}
\ee
where $\sigma^2=\case{1}/{2}\sigma_{ab}\sigma^{ab}$ and
$\omega^2=\case{1}/{2}\omega_{ab}\omega^{ab}$ are the shear
and vorticity magnitudes and $\mu$ and $p$ are the energy density
and pressure respectively.
%%%%%%%%%%%%%%%%%%%%%%%%%%%%%%%%%%%%%%%%%%%%%%%%%
\subsection{Matter and radiation}
%%%%%%%%%%%%%%%%%%%%%%%%%%%%%%%%%%%%%%%%%%%%%%%%%
Fixing $u^a$ so that it corresponds to the Landau\hs Lifshitz (energy) 
frame \cite{bi:LL}\footnote{In this frame the total energy 
flux $q_a$ vanishes.} and considering only small deviations 
from equilibrium (so that the velocities of the matter and radiation 
relative to this frame are small), the total energy momentum tensor 
for matter $(m)$ and radiation $(r)$ is given by
\be
T_{ab}=\mu u_au_b+ph_{ab}+\pi^{(r)}_{ab}\;,
\label{eq:EMT}
\ee 
where
\be 
\mu=\mu_{(m)}+\mu_{(r)}\;,~~~p=\case{1}/{3}\mu_{(r)}\;,
\ee
and $\pi^{(r)}_{ab}$ is the anisotropic pressure of the radiation.
%%%%%%%%%%%%%%%%%%%%%%%%%%%%%%%%%%%%%%%%%%%%%%%%%%%%%%%%%%
\subsection{Component fluid equations}
%%%%%%%%%%%%%%%%%%%%%%%%%%%%%%%%%%%%%%%%%%%%%%%%%%%%%%%%%%
Relative to this frame the conservation of energy and momentum for
non\hs interacting matter and radiation are given by \cite{bi:DBE}
\be
\dot{\mu}_{(r)}+\case{4}/{3}\mu_{(r)}\Theta+\n^aq^{(r)}_a
+2q^{(r)}_a\dot{u}^a+\pi^{(r)}_{ab}\sigma^{ab}=0\;,
\ee
\label{eq:energy}
\be
\dot{\mu}_{(m)}+\mu_{(m)}\Theta+\n^aq^{(m)}_a=0\;,
\ee
and
\ber
\lefteqn{
h^c{}_a\dot{q}^{(r)}_c+\case{4}/{3}\mu_{(r)}\dot{u}_a
+\case{1}/{3}\n_a\mu_{(r)}+\n^b\pi^{(r)}_{ab}}\\
&&+\dot{u}^b\pi^{(r)}_{ab}+\left(\sigma_{ab}+w_{ab}
+\case{4}/{3}\Theta h_{ab}\right)q^b_{(r)}=0\;,
\ear
\be
h^c{}_a\dot{q}^{(m)}_c+\mu_{(m)}\dot{u}_a+\left(\sigma_{ab}+\omega_{ab}
+\case{4}/{3}\Theta h_{ab}\right)q^b_{(m)}=0\;,
\ee
where
\be
q^{(r)}_a=\case{4}/{3}\mu_{(r)}V^{(r)}_a\;,
~~q^{(m)}_a=\mu_{(m)}V^{(m)}_a\;,~~q^{(r)}_a+q^{(m)}_a=0\;,
\ee
and $V^{(r)}_a$ and $V^{(m)}_a$ are the velocities of the matter 
and radiation relative to $u^a$:
\be
V^{(r)}_a=u^{(r)}_a-u_a\;,~~~~~~V^{(m)}_a=u^{(m)}_a-u_a\;. 
\ee
%%%%%%%%%%%%%%%%%%%%%%%%%%%%%%%%%%%%%%%%%%%%%%%%%%%%%%
\subsection{Total fluid equations}
%%%%%%%%%%%%%%%%%%%%%%%%%%%%%%%%%%%%%%%%%%%%%%%%%%%%%%
Because we have chosen to work in the {\it energy frame}, the 
conservation equations for the total fluid are considerably simpler 
than those for the individual components:
\be
\dot{\mu}+h\Theta+\pi^{(r)}_{ab}\sigma^{ab}=0\;,
\ee
\be
h\dot{u}_a+\n_ap+\n^b\pi^{(r)}_{ab}+\dot{u}^b\pi^{(r)}_{ab}=0\;,
\ee
where
\be
h=\mu_{(m)}+\case{4}/{3}\mu_{(r)}\;,
\ee
is the sum of the total energy density and pressure.
%%%%%%%%%%%%%%%%%%%%%%%%%%%%%%%%%%%%%%%%%%%%%%%%%%%%%%%
\subsection{FRW models}
%%%%%%%%%%%%%%%%%%%%%%%%%%%%%%%%%%%%%%%%%%%%%%%%%%%%%%%
In the case of a FRW universe, $u^a_{(m)}=u^a_{(r)}=u^a$ and 
$\pi^{(r)}_{ab}=0$, so the energy momentum tensor (\ref{eq:EMT})
necessarily reduces to the perfect fluid form 
\be
T_{ab}=\mu u_au_b+ph_{ab},
\ee
and
\be 
\dot{u}_a=\n_a p=0\;,
\ee
so dynamics of matter\hs radiation models are completely 
determined by the energy conservation equations
\be
\dot{\mu}_{(r)}+4\mu_{(r)}\Theta=0\;,
\ee
\be
\dot{\mu}_{(m)}+3\mu_{(m)}\Theta=0\;,
\label{eq:back1}
\ee
\be
\dot{\mu}+h\Theta=0\;,
\ee
together with the Friedmann equations
\be
H^2+\frac{K}{a^2}=\case{1}/{3}\kappa\mu\;,
\ee
where $K$ is the spatial curvature constant, and
\be
3\dot{H}+3H^2+\case{1}/{2}\kappa\left(\mu+p\right)=0\;,
\label{eq:back2}
\ee 
which is (\ref{eq:theta}) specialized to a FRW model.
%%%%%%%%%%%%%%%%%%%%%%%%%%%%%%%%%%%%%%%%%%%%%%%
\section{A gauge\hs invariant measure of CMB temperature anisotropies}
%%%%%%%%%%%%%%%%%%%%%%%%%%%%%%%%%%%%%%%%%%%%%%%
\begin{figure}
\epsffile{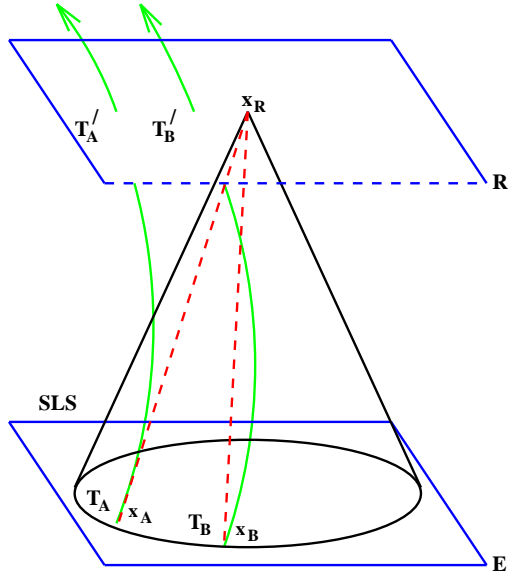}
\caption{A diagram illustrating the geometry of the Sachs\hs Wolfe effect.
It clearly shows that the observed temperature anisotropy at ${\bf x}_R$
is given by the difference in temperature of the CMB observed in different
directions on the plane of the sky, corresponding to different points on the 
surface of last scattering (SLS).}
\end{figure}
The aim of this paper is to derive in the simplest possible way a 
covariant and gauge\hs invariant formula that describes the observed CMB 
temperature anisotropy $\Delta T_R/\bar{T}_R$. This improves on earlier 
work by Russ {\it et al} \cite{bi:russetal}, where a similar formula 
was obtained by taking first order variations of the redshift.  

Before sketching this derivation, let us first clarify the difference 
between the gauge\hs invariant temperature perturbation $\delta T_R/\bar{T}_R$ 
and the observed temperature anisotropy, since this continues to be a 
source of considerable confusion in the literature. 

If the point of observation is defined by the spacetime point 
${\bf x}_R$, with coordinates $x^a_R$, then the temperature measured 
by an observer at ${\bf x}_R$ in a direction ${\bf e}_A$, with components
$e^a_A$, can be decomposed into the average bolometric temperature 
$\bar{T}_R(x^a_R)$ and the covariant and gauge\hs invariant 
temperature $\delta T_R(x^a,e^a_A)$ \cite{bi:MES}:
\ber
T_R({\bf x}_R,{\bf e}_A)&=&\bar{T}_R({\bf x}_R)
+\delta T_R({\bf x}_R,{\bf e}_A)\;,\label{eq:deftemp} \\
\bar{T}_R(x^a_R)&=&\frac{1}{4\pi}\int_{4\pi}T_R(x^a_R,e^a_A)d\Omega\;.
\ear
The observed temperature anisotropy is given by 
\be
\frac{\Delta T_R}{\bar{T}_R}=\frac{T_R({\bf x}_R,{\bf e}_A)
-T_R({\bf x}_R,{\bf e}_B)}{\bar{T}_R({\bf x}_R)}\;,
\ee
where ${\bf e_A}$ and ${\bf e}_B$ correspond to two different 
directions on the plane of the sky corresponding to two different points
of emission ${\bf x}_A$ and ${\bf x}_B$ on the last scattering 
surface (see figure 1 above). If $\delta T_R\ll 1$ this becomes
\be
\frac{\Delta T_R}{\bar{T}_R}
=\frac{\delta T_R}{\bar{T}_R}({\bf x}_R,{\bf e}_A)
-\frac{\delta T_R}{\bar{T}_R}({\bf x}_R,{\bf e}_B)\;,
\label{eq:aniso}
\ee
since the average bolometric temperature contributions to $T$ cancel out 
when subtracted at the point of observation ${\bf x}_R$.
%%%%%%%%%%%%%%%%%%%%%%%%%%%%%%%%%%%%%%%%%%%%%%%
\section{A gauge\hs invariant formula for $\delta T_R/\bar{T}_R$}
%%%%%%%%%%%%%%%%%%%%%%%%%%%%%%%%%%%%%%%%%%%%%%%
Having clarified the  meaning of temperature perturbations and the 
observed temperature anisotropy, we  will now proceed to derive from first
principles an expression for $\delta T_R/\bar{T}_R$ in terms of the 
covariant quantities defined in section \ref{sec:basic}.

In principle what one should do is integrate the Louville
equation in curved spacetime for a generalized gauge\hs invariant 
distribution function \cite{bi:SME,bi:MES} first through the decoupling 
phase and then from decoupling until today to obtain the perturbed 
spectrum of photon energies observed in a given direction ${\bf e}_A$ on the 
sky. The temperature anisotropy is then obtained 
by subtracting these results for two independent directions.

It turns out however, that because photons are essentially collisionless 
after last scattering, the complete physical content of the Louville 
equation is contained in the geodesic equation, so it is much simpler 
instead to integrate the photon energies $E$ up null geodesics 
connecting points of emission on the last scattering surface with the point 
of observation ${\bf x}_R$ here and now.

Photons moves on null geodesics $x^a(\lambda)$ where $\lambda$ 
is an affine parameter. The null vector to these geodesics is
\be 
p^a=\frac{dx^a}{d\lambda}\;,~~~~p^ap_a=0\;,
\ee 
and satisfies the geodesic equation 
\be p^a{}_{;b}p^b=0\;.
\label{eq:geodesic}
\ee 
The $3+1$ decomposition of this null vector 
\be 
p^a=E(u^a+e^a)\;,~~~e^au_a=0\;,~~~e^ae_a=1\;,
\ee 
defines the energy
\be 
E=-p^au_a
\label{eq:ehlers}
\ee 
of a photon relative to the four\hs velocity of the observer $u^a$ and 
its relative (spatial) direction of motion $e^a$. 

The temperature $T_R=T_R({\bf x}_R,{\bf e}_E)$ of the CMB observed 
at the reception point $R$ with spacetime coordinates $x^a_R$ in a 
given direction ${\bf e}_E$ is given by
\be 
\frac{T_R}{T_A}=\frac{1}{1+z}=\frac{(p^au_a)_R}{(p^au_a)_A}\;,
\label{eq:temp}
\ee 
where $T_A=T_A({\bf x}_A)$ is the temperature at the emission point 
${\bf x}_A$ and $z=z({\bf e}_A)$ is the redshift between emission 
and reception. 

The equation describing the variation of a photon's energy along
a null geodesic (parameterized by $\lambda$) is obtained by 
differentiating equation (\ref{eq:ehlers}) with respect to 
$\lambda$ and using the  geodesic equation (\ref{eq:geodesic}). This
yields
\be
\frac{dE}{d\lambda}=-u_{a;b}p^ap^b\;.
\ee
Substituting for $u_{a;b}$ from equations (\ref{eq:uab}) and (\ref{eq:cov}) 
gives
\be
\frac{dE}{d\lambda}=-\case{1}/{3}\Theta E^2
-\sigma_{ab}p^ap^b-E\dot{u}_ap^a\;.
\label{eq:null1}
\ee
At this point one could already integrate for $E$, however for the 
problem that we wish to discuss, it is more convenient to substitute for 
the expansion $\Theta$ in terms of known gauge\hs invariant perturbation 
variables. This is best done by projecting the spatial gradient of the 
radiation energy density $X^{(r)}_a=\n_a\mu_{(r)}$ along the null 
vector $p^a$:
\be
p^aX^{(r)}_a=p^a\frac{\p\mu_{(r)}}{\p x^a}+\dot{\mu}_{(r)}p^au_a
=\frac{d\mu_{(r)}}{d\lambda}-E\dot{\mu}_{(r)}\;.
\ee
Substituting for $\dot{\mu}_{(r)}$ from the energy conservation equation for 
the radiation (\ref{eq:energy}) one obtains
\be
p^aX^{(r)}_a=\frac{d\mu_{(r)}}{d\lambda}+\case{4}/{3}\mu_{(r)}E\Theta
+E\pi^{(r)}_{ab}\sigma^{ab}+E\n^aq^{(r)}_a+2E\dot{u}^aq^{(r)}_a\;.
\ee
It follows that
\be
E\Theta=\frac{3}{4\mu_{(r)}}\left[p^aX^{(r)}_a-\frac{d\mu_{(r)}}{d\lambda}
-E\pi^{(r)}_{ab}\sigma^{ab}-E\n^aq^{(r)}_a-2E\dot{u}^aq^{(r)}_a\right]\;,
\ee
and substituting this into equation (\ref{eq:null1}) gives
a completely general equation for the variation of a photons'
energy along a null geodesic:
\be
\frac{1}{E}\frac{dE}{d\lambda}-\frac{1}{4\mu_{(r)}}
\frac{d\mu_{(r)}}{d\lambda}=-F\;,
\ee
where $F$ is given by
\ber
F&=&\frac{1}{4\mu_{(r)}}X^{(r)}_ap^a-\frac{E}{4\mu_{(r)}}
\left[\pi^{(r)}_{ab}\sigma^{ab}+\n^aq^{(r)}_a
+2\dot{u}^aq^{(r)}_a\right]\nonumber\\
&+&\frac{1}{E}\sigma_{ab}p^ap^b+\dot{u}_ap^a\;.
\label{eq:func}
\ear
Integrating up a null geodesic from the point of emission ${\bf x}_A$
at last scattering to the point of reception ${\bf x}_R$ and using
equation (\ref{eq:temp}) one obtains an exact formula for
the temperature at reception $T_R$:
\be
\ln\left(\frac{T_R}{T_A}\right)=\frac{1}{4}\ln
\left(\frac{\mu_{(r)R}}{\mu_{(r)A}}\right)-\int^R_AFd\lambda\;.
\ee
Substituting for $T_R$ from equation (\ref{eq:deftemp}) and using 
the Stephan\hs Boltzmann law $\mu_{(r)}=aT^4$ we obtain
\be
\ln\left[1+\frac{\delta T_R}{\bar{T}_R}\right]=-\int^R_AFd\lambda\;.
\label{eq:exact}
\ee
It is important to realize that apart from assuming that photons are 
collision free, no approximations have been made in this section so far, 
and the above result is therefore valid for any choice of background 
geometry and matter description\footnote{In the case of a  
FRW model, $F=0$ and $\mu_{(r)}\propto a^{-4}$ so the standard 
result of $T_R/T_A=a_A/a_R$ is recovered.}.
%%%%%%%%%%%%%%%%%%%%%%%%%%%%%%%%%%%%%%%%%%%%%%%
\subsection{Linearization about FRW models} \label{sec:lin}
%%%%%%%%%%%%%%%%%%%%%%%%%%%%%%%%%%%%%%%%%%%%%%%
To obtain an expression for $\delta T_R/\bar{T}_R$
valid in  an {\it almost} FRW model (a spacetime where these 
variables  are small) we approach this universe from the equations
valid in a general spacetime rather that adopting the standard 
procedure of perturbing an exact FRW model. The linearization 
procedure we apply consists of dropping terms such as 
$\pi^{(r)}_{ab}\sigma^{ab}$ in equation (\ref{eq:func}), i.e. terms 
which are second order in the gauge\hs invariant variables, 
retaining only those terms which are linear, for 
example $X^{(r)}_a$ \cite{bi:EB}. Linearizing equation (\ref{eq:exact}) 
following this procedure yields the following result
\be
\frac{\delta T_R}{\bar{T}_R}=-\int^R_A\left(\frac{1}{4a}
\dd^{(r)}_ap^a-\frac{1}{3a}\mu_{(r)}\n^aV^{(r)}_a+a\sigma_{ab}p^ap^b
+\dot{u}_ap^a\right)d\lambda
\label{eq:lin}
\ee
where $\dd^{(r)}_a\equiv aX^{(r)}_a/\mu_{(r)}$ and we have used the 
normalized background (FRW) expression for the photon energy $E=1/a$. 

This formula expresses the generation of CMB anisotropies by cosmological
perturbations in it's clearest form with each term having a direct physical
interpretation. First it should be noted that $\dd^{(r)}_a$, $V^{(r)}_a$ and 
$\dot{u}_a$ contain both a scalar and vector part, while the shear 
$\sigma_{ab}$ is made up of contributions due to scalar, vector and tensor 
perturbations. Focusing on scalar perturbations, $\dd^{(r)}_a$ and 
$V^{(r)}_a$ characterize density and velocity perturbations in the 
radiation relative to $u^a$, the acceleration term $\dot{u}^a$ represents
possible pressure suppression effects \cite{bi:SEX} and the shear 
relates to perturbations in the gravitational potential.

We can express this result in terms of total matter variables by using
the following results \cite{bi:DBE}:
\be
\frac{1}{4}\dd^{(r)}_a+a\dot{u}_a=\frac{1}{3h}\left(1-3c^2_s\right)\mu
\dd_a+\frac{1}{3}\frac{\mu^2_{(m)}}{h^2}S^{(rm)}_a\;,
\ee
\be
V^{(r)}_a=\frac{\mu_{(m)}}{h}V^{(rm)}_a\;,
\ee
where 
\be
c^2_s=\frac{4\mu_{(r)}}{3(4\mu_{(r)}+3\mu_{(m)})}
\ee
is the speed of sound in the total fluid,
\be
\mu\dd_a=\mu_{(m)}\dd^{(m)}_a+\mu_{(r)}\dd^{(r)}_a
\ee
is the total perturbation in the energy density and
\be
S^{(rm)}_a=\frac{1}{4}\dd^{(r)}_a-\frac{1}{3}\dd^{(m)}_a\;,~~
V^{(rm)}_a=u^{(r)}_a-u^{(m)}_a
\ee
are the entropy and relative velocity perturbations 
respectively \cite{bi:DBE}.

At the time of decoupling, if the present value of the density parameter 
$\Omega_0>0.1$, the universe is matter dominated to a good approximation, 
so $h\rightarrow\mu_{(m)}$ and $c^2_s\rightarrow 0$. It follows that  
above results reduce to 
\be 
\frac{1}{4}\dd^{(r)}_a+a\dot{u}_a=\frac{1}{3}\dd_a
+\frac{1}{3}S^{(rm)}_a\;,~~~V^{(r)}_a=V^{(rm)}_a
\label{eq:velocity}
\ee
and the expression for $\delta T_R/\bar{T}_R$ becomes
\be
\frac{\delta T_R}{\bar{T}_R}={\cal A}
-\int^R_A\left(\frac{1}{3a}\dd_ap^a+a\sigma_{ab}p^ap^b\right)d\lambda\;,
\label{eq:main}
\ee
where
\be
{\cal A}=-\int^R_A\frac{1}{3a}\left(S^{(rm)}_ap^a
-\n^aV^{(rm)}_a\right)d\lambda\;.
\label{eq:nonadiab}
\ee
We thus have two contributions: one due to perturbations in the total
energy density and pressure and the other, ${\cal A}$, due to the difference
in the dynamical behavior of the matter and radiation density and velocity
perturbations.
%%%%%%%%%%%%%%%%%%%%%%%%%%%%%%%%%%%%%%%%%%%%%%%
\section{The temperature anisotropy due to
gravitational potential perturbations}\label{sec:Newt}
%%%%%%%%%%%%%%%%%%%%%%%%%%%%%%%%%%%%%%%%%%%%%%%
In this section we will deal with the contribution to CMB anisotropies due to 
gravitational potential fluctuations. In order to do this it is first 
convenient to write the formula (\ref{eq:main}) in terms of the electric 
$E_{ab}$ and magnetic $H_{ab}$ parts of the Weyl tensor. This is done 
by using the two linearized Bianchi identities which relate to 
$E_{ab}$ \cite{bi:BDE}:
\be
\dot{E}_{ab}+3HE_{ab}+h^f{}_{(a}\eta_{b)cde}u^cH_f{}^{d;e}
+\case{1}/{2}h\sigma_{ab}=0\;,
\label{eq:Edot}
\ee
\be
a\n^aE_{ab}=\case{1}/{3}\kappa\mu\dd_a\;,
\label{eq:divE}
\ee
to substitute for the shear and $\dd_a$ in (\ref{eq:main}). This leads 
straightforwardly to the following result:
\ber
\frac{\delta T_R}{\bar{T}_R}&=&{\cal A}
-\int^R_A\frac{1}{\mu_{(m)}}\left[\n^aE_{ab}p^b
-2a\left(\dot{E}_{ab}+3HE_{ab}\right)p^ap^b\right.\nonumber\\
&+&\left. h^f{}_{(a}\eta_{b)cde}u^cH_f{}^{d;e}p^ap^b\right]d\lambda\;.
\label{eq:main2}
\ear
This expression is closely related the formulae derived by Magueijo 
\cite{bi:magueijo} and Durrer \cite{bi:durrer1,bi:durrer2}.

In the case of scalar gravitational potential fluctuations the magnetic 
part of the Weyl tensor $H_{ab}$ vanishes (since it only contributes to 
vector and tensor perturbations \cite{bi:BDE,bi:goode}), 
so (\ref{eq:main2}) reduces to 
\be
\frac{\delta T_R}{\bar{T}_R}={\cal A}
-\int^R_A\frac{1}{\mu_{(m)}}\left[\n^aE_{ab}p^b
-2a\left(\dot{E}_{ab}+3HE_{ab}\right)p^ap^b\right]d\lambda\;.
\label{eq:main3}
\ee
For scalar perturbations, the electric part of the Weyl tensor (with
wave number $k$) is related to the harmonic component of the 
perturbed gravitational potential $\Phi_k=\Phi_k(t)$ as 
follows \cite{bi:BDE}:
\be
E_{ab}=\frac{k^2}{a^2}\Phi_kQ_{ab}\;,
\label{eq:electric}
\ee
where $Q_{ab}$ is a covariantly defined scalar harmonic and $k$ is the 
eigenvalue associated with $Q_{ab}$ (it is the wave number if the background 
FRW spacetime is flat). A discussion of these harmonics and their 
properties is given in appendix A and \cite{bi:BDE}. 

Substituting (\ref{eq:electric}) into (\ref{eq:main3}) and using equations
(\ref{eq:harm2}) and (\ref{eq:harm4}) in appendix A, we obtain
\ber
\frac{\delta T_R}{\bar{T}_R}&=&{\cal A}
-2\int^R_A\frac{1}{a^3\mu_{(m)}}\left[\frac{1}{3}\left(3K-k^2\right)
\left(a\Phi_kQ\right)'\right.\nonumber\\
&-&\left.K\left(a\Phi_k\right)Q'
-a^3\left(a\Phi_k\right)'\left(aQ'\right)'\right]d\lambda\;,
\label{eq:main4}
\ear
where the prime denotes differentiation with respect to $\lambda$.
In the background FRW model the energy conservation equation (\ref{eq:back1})
can be integrated to give 
\be
\mu_{(m)}=\alpha a^{-3}\;,~~~\alpha=\mu_Ea_E^3\;,
\label{eq:hback}
\ee
where $\mu_E$ and $a_E$ are the background values for the energy density and 
scale factor at the time of emission. 

Substituting (\ref{eq:hback}) into (\ref{eq:main4}) and integrating the 
first term by parts gives
\ber
\frac{\delta T_R}{\bar{T}_R}&=& {\cal A}+\case{2}/{3}
\left(3K-k^2\right)\left(\Phi_kQ\right)_E\nonumber\\
&+&\frac{2}{\alpha}\int^R_A\left[KQ+a^3\left(aQ'\right)'\right]
\left(H\Phi_k+\dot{\Phi}_k\right)d\lambda\;,
\label{eq:main5}
\ear
where we have dropped the term evaluated at reception since it has no angular 
dependence. This result is true for a general FRW background. The 
$\dot{\Phi}_k$ represents the integrated or Rees\hs Sciama effect which is 
important if the potential is non\hs stationary, for example in open 
universe models. 
%%%%%%%%%%%%%%%%%%%%%%%%%%%%%%%%%%%%%%%%%%%%%%%%%%%%
\subsection{Large scale temperature anisotropies}
%%%%%%%%%%%%%%%%%%%%%%%%%%%%%%%%%%%%%%%%%%%%%%%%%%%%
To further simplify the above problem, we will now make the standard
assumptions of assuming that the background is a flat ($K=0$) FRW model, 
and consider temperature anisotropies arising as a result of density 
perturbations on scales much larger than the Hubble radius at decoupling. 
In this case the Friedmann and energy conservation equations 
(\ref{eq:back1}\hs\ref{eq:back2}) lead to the following background 
evolution for the scale factor:
\be
a=\left(\beta t\right)^{\case{2}/{3}}\;,~~~\beta^2=\case{3}/{4}\alpha\;,
\label{eq:back3}
\ee
and in the matter dominated regime the potential fluctuations $\Phi_k$ 
satisfy the following differential equation \cite{bi:EHB}:  
\be
\ddot{\Phi}_k+4H\dot{\Phi}_k=0\;,
\label{eq:pert}
\ee
which follows from (\ref{eq:Edot}) and the shear propagation equation when
$H_{ab}=0$. Substituting for the scale factor (\ref{eq:back3}) 
in (\ref{eq:main5}) and integrating by parts, 
using (\ref{eq:pert}) to substitute for second derivatives 
in $\Phi_N$ gives the following result:
\ber
\frac{\delta T_R}{\bar{T}_R}&=&{\cal A}
+\left(\Phi_kQ\right)_A+\case{2}/{3}\Hs_A
\left(a\Phi_kQ_ap^a\right)_A
-\case{2}/{9}\Hs_A^2\left(\Phi_kQ\right)_A\nonumber\\
&+&\case{2}/{3}\Hs_A\left(\case{\dot{a}}/{a}\right)^{-1}
\left(a\dot{\Phi}_kQ_ap^a\right)_A+\int^R_A\dot{\Phi}_kQdt\;.
\ear
where
\be
\Hs_A=\frac{k}{a_AH_A}=\left(\frac{\lambda_H}{\lambda}\right)_A
\label{eq:calH}
\ee
is the ratio of the Hubble scale $\lambda_H=1/H$ to the comoving scale 
$\lambda$ at the time of decoupling and the last term represents
the integrated Sachs\hs Wolfe effect.

If the gravitational potential $\Phi_k$ is approximately constant
and we consider scales much larger than the Hubble radius at decoupling, 
so that $\Hs_A\ll 1$, we recover the well known result of Sachs and Wolfe, 
together with additional contributions due to velocity and entropy 
perturbations
\be
\frac{\delta T_R}{\bar{T}_R}={\cal A}+\left(\Phi_kQ\right)_A\;.
\ee
Finally if we subtract this result for two independent directions 
($A$ and $B$), the observed temperature anisotropy 
$\Delta T_R/\bar{T}_R$ is obtained:
\be
\frac{\Delta T_R}{\bar{T}_R}=\Delta{\cal A}+\Delta\left(\Phi_kQ\right)\;,
\label{eq:SW}
\ee
where
\be
\Delta\left(\Phi_kQ\right)=\left(\Phi_kQ\right)_A-\left(\Phi_kQ\right)_B\;.
\ee
is the difference in the gravitational potential between separate points
$A$ and $B$ on the SLS. 
%%%%%%%%%%%%%%%%%%%%%%%%%%%%%%%%%%%%%%%%%%%%%%%%%%%%
\section{The adiabatic assumption} \label{sec:adiabatic}
%%%%%%%%%%%%%%%%%%%%%%%%%%%%%%%%%%%%%%%%%%%%%%%%%%%%
One of the most common assumptions made when discussing large scale CMB 
anisotropies is that the perturbations in the total energy density are 
adiabatic at the time of decoupling. Let us now consider whether or not 
this assumption is consistent and how it affects the results presented 
above. In order to achieve clarity on this issue, we need to 
consider (i) the large scale evolution  of density, entropy and 
relative velocity perturbations during the collision dominated period 
prior to decoupling as these provide the initial conditions at the time 
of decoupling; (ii) how these initial conditions relate to 
the correct placing of the surface of last scattering 
and (iii) the evolution of these perturbations after decoupling.  
%%%%%%%%%%%%%%%%%%%%%%%%%%%%%%%%%%%%%%%%%%%%%%%%%%%%%
\subsection{Before decoupling}
%%%%%%%%%%%%%%%%%%%%%%%%%%%%%%%%%%%%%%%%%%%%%%%%%%%%%
On large scales, in the matter dominated limit, the dynamics of 
density $\Delta_{(r)}\equiv a\n^a\dd^{(r)}_a$, entropy 
$S_{(rm)}\equiv a\n^aS^{(rm)}_a$ and relative 
velocity $V_{(rm)}\equiv a\n^aV^{(rm)}_a$ perturbations in 
a photon\hs baryon universe are described by the following set of 
equations \cite{bi:DBE}:
\be
\ddot{\Delta}_{(r)}+2H\dot{\Delta}_{(r)}-\frac{1}{2}h\Delta_{(r)}
=-\frac{4}{3}\left[\frac{1}{2}hS_{(rm)}
-H\left(1-\frac{h}{\mu_{(m)}}R_c\right)\dot{S}_{(rm)}\right]\;,
\label{eq:ode1}
\ee
\be
\ddot{S}_{(rm)}+H\left(\frac{4}{3}\frac{\mu_{(r)}}{h}
+\frac{h}{\mu_{(m)}}R_c+1\right)\dot{S}_{(rm)}=0\;,
\label{eq:ode2}
\ee
and
\be
\dot{V}_{(rm)}+H\left(\frac{4}{3}\frac{\mu_{(r)}}{h}
+\frac{h}{\mu_{(m)}}R_c\right)V_{(rm)}=-\frac{1}{4}\frac{1}{ha^4}
\Delta_{(r)}\;,
\label{eq:ode3}
\ee
where $R_c(t)=1/H\tau_c$ is the ratio of the horizon size to the mean free 
path for photons colliding with electrons and $\tau_c$ is the mean 
collision time of photons with electrons.

Equations (\ref{eq:ode1}\hs\ref{eq:ode3}) can be integrated to give
solutions for density, entropy and velocity perturbations 
in the tightly coupled regime before decoupling:
\be
\Delta_{(r)}=t^{2/3}A_a\;,
\ee
\be
S_{(rm)}=S_0+B\int^{t_{dec}}_0\frac{1}{a}e^{-P(t)}dt\;,
\label{eq:sol2}
\ee
\be
V_{(rm)}=C(t)e^{-P(t)}\;.
\label{eq:sol3}
\ee
Before decoupling $R_c(t)$ is much greater than unity since the mean collision 
time between photons and baryons tends to zero, and therefore $P(t)\gg 1$
in this limit. This means that the solution for $S_{(rm)}$ (\ref{eq:sol2})
settles down to a constant value immediately after it is provoked, and so 
{\it entropy perturbations have essentially one mode which is constant 
in time}. This is due to the fact that the matter and radiation are
so tightly coupled that the matter cannot move relative to the 
radiation. This behavior can be seen by looking at the solution for
$V_{(rm)}$ (\ref{eq:sol3}) which is exponentially driven to zero. 
These solutions imply that if the perturbations are initially adiabatic,
as suggested by many inflationary scenarios, they will remain so until 
the time of decoupling. 
%%%%%%%%%%%%%%%%%%%%%%%%%%%%%%%%%%%%%%%%%%%%%%%%%%%%%
\subsection{Adiabatic perturbations at decoupling}
%%%%%%%%%%%%%%%%%%%%%%%%%%%%%%%%%%%%%%%%%%%%%%%%%%%%%
Given that the pre\hs decoupling perturbation dynamics can lead to adiabatic
initial conditions, let us consider whether they are compatible with the 
proper placing of the SLS \cite{bi:panek,bi:SXEK,bi:SES,bi:adiab1}. 

The correct way of placing this surface is by determining where the optical
depth due to Thompson scattering is unity. This occurs, to first order,
where the radiation temperature, which is equal to the matter temperature
in the strongly coupled region prior to decoupling: $T_{(r)}=T_{(m)}=T$,
reaches the matter ionization temperature, so the last scattering
event $A$ on each null geodesic is characterized by
\be
T_{A}=T_{ion}\;.
\ee
Thus if we take decoupling as happening essentially instantaneous, the SLS
is, to good approximation, a surface of constant radiation temperature and
so by the Stefan\hs Boltzmann law $\mu_{(r)}=aT^4$, also one of constant 
radiation density:
\be
\Delta T=0~~\Rightarrow~~\tilde{\dd}^{(r)}_a=0\;,
\label{eq:constT}
\ee
where $\Delta T=T_A-T_B$ is the difference in temperature between separate 
points of emission $x_A$ and $x_B$ on the SLS (see figure 1) and 
$\tilde{\dd}^{(r)}_a$ is the spatial variation of $\mu_{(r)}$ 
orthogonal to the normals $n^a$ to the surfaces of constant radiation 
density ($\tilde{\dd}^{(r)}_an^a=0$). Transforming to the energy frame $u^a$,
we can relate $\tilde{\dd}^{(r)}_a$ to $\dd^{(r)}_a$:
\be
\tilde{\dd}^{(r)}_a=\dd^{(r)}_a+4aHV_a\;,~~V^a=u^a-n^a\;,
\label{eq:grad1}
\ee
and taking its spatial divergence we obtain the 
corresponding result for scalar perturbations:
\be
\tilde{\Delta}_{(r)}=\Delta_{(r)}+4aHV\;,
\label{eq:grad2}
\ee
where
\be 
\tilde{\Delta}_{(r)}\equiv a\n^a\dd^{(r)}_a\;,~~~V\equiv a\n^aV_a\;.
\ee
If the perturbations are adiabatic at decoupling:
\be
S_{(rm)}=0\Rightarrow \Delta_{(r)}=\frac{4}{3}\Delta_{(m)}\;,~~V_{(rm)}=0\;,
\label{eq:grad3}
\ee
so equation (\ref{eq:grad2}) becomes 
\be
\tilde{\Delta}_{(r)}=\frac{4}{3}\Delta_{(m)}+4aHV\;.
\ee
Hence, for a surface of constant radiation density $\tilde{\Delta}_{(r)}=0$, 
which defines the SLS, we find that:
\be
\Delta_{(m)}=-3HaV\;.
\label{eq:vel}
\ee
Using equations (\ref{eq:divE}) and we can relate $\Delta_{(m)}$ 
to the electric part of the Weyl tensor:
\be
a^2\n^a\n^bE_{ab}=\case{1}/{3}\kappa\mu\Delta_{(m)}\;,
\ee
and combining this with (\ref{eq:vel}) we obtain: 
\be
\n^a\n^bE_{ab}=-\case{\mu H}/{a}V\;.
\label{eq:vel2}
\ee
Using (\ref{eq:electric}) and the results in Appendix A, 
the LHS of (\ref{eq:vel2}) can be written in terms of the 
harmonic components $\Phi_k$ of the perturbed gravitational 
potential, while the RHS can be decomposed into its harmonic
components $V_k$ (see equation 102 in \cite{bi:DBE}):
\be
\n^a\n^bE_{ab}=\case{2}/{3}\case{k^4}/{a^4}\left(\Phi_kQ\right)\;,~~~
V=-kV_kQ\;.
\ee 
Substituting these results into (\ref{eq:vel2}) and using 
equation (\ref{eq:calH}) gives the following result:
\be
\Phi_kQ=\case{3}/{2}{\cal H}^{-3}V_kQ\;.
\ee
It therefore follows that for adiabatic perturbations 
the large\hs scale temperature anisotropy 
\be
\frac{\Delta T_R}{\bar{T}_R}=\Delta\left(\Phi_kQ\right)
=\case{3}/{2}{\cal H}^{-3}_A\Delta\left(V_kQ\right)
\ee
is simply related to the motion of the matter relative to the 
surfaces of constant temperature:
%%%%%%%%%%%%%%%%%%%%%%%%%%%%%%%%%%%%%%%%%%%%%%%%%
\subsection{Adiabatic perturbations after decoupling}
%%%%%%%%%%%%%%%%%%%%%%%%%%%%%%%%%%%%%%%%%%%%%%%%%
After decoupling, in the free propagating domain, $R_c(t)\ll 1$, so the 
large\hs scale perturbation equations (\ref{eq:ode1}\hs\ref{eq:ode3}) 
reduce to
\be
\ddot{\Delta}_{(r)}+2H\dot{\Delta}_{(r)}-\frac{1}{2}h\Delta_{(r)}
=-\frac{4}{3}\left[\frac{1}{2}hS_{(rm)}-H\dot{S}_{(rm)}\right]\;,
\label{eq:ode4}
\ee
\be
\ddot{S}_{(rm)}+H\left(\frac{4}{3}\frac{\mu_{(r)}}{h}
+1\right)\dot{S}_{(rm)}=0\;,
\label{eq:ode5}
\ee
\be
\dot{V}_{(rm)}+\frac{4}{3}\frac{\mu_{(r)}H}{h}V_{(rm)}
=-\frac{1}{4}\frac{1}{ha^4}\Delta_{(r)}\;,
\label{eq:ode6}
\ee
and in the matter dominated limit they can be integrated to give the 
following solutions
\be
\Delta_{(r)}=At^{2/3}\;,
\label{eq:sol4}
\ee
\be
S_{(rm)}=S_0+Bt^{1/3}\;,
\label{eq:sol5}
\ee
\be
V_{(rm)}=Ct\;.
\label{eq:sol6}
\ee
These solutions demonstrate that after decoupling generic density 
perturbations do not satisfy the adiabatic condition 
$S_{(rm)}=V_{(rm)}=0$. This is due to the fact that the average 
velocity of the radiation does not proceed along geodesics, 
while the matter does. Thus any perturbation that starts off 
adiabatic at last scattering will not remain so. 
%%%%%%%%%%%%%%%%%%%%%%%%%%%%%%%%%%%%%%%%%%%%%%%%%
\section{Conclusion} \label{sec:end}
%%%%%%%%%%%%%%%%%%%%%%%%%%%%%%%%%%%%%%%%%%%%%%%%%
In this paper we have calculated a fully covariant formula for 
the CMB temperature anisotropy improving on earlier work
by Russ {\it et al} \cite{bi:russetal}. This formulation has a 
number of distinct advantages over the more standard approaches
as it is independent of gauge conditions, non\hs local splittings
of spacetime, and related Fourier decompositions of perturbations
around a FRW metric. Furthermore the results are relatively simple
and easy to interpret. For scalar perturbations
we recovered the dominant Sachs\hs Wolfe term, together with 
the Rees\hs Sciama effect which contributes to large scale 
CMB anisotropy only if the perturbations to the gravitational 
potential are non\hs stationary. 

We also examined the validity of the assumption that the density 
perturbations are adiabatic at decoupling and showed that if the 
surface of last scattering is correctly placed and the background 
is assumed to be a flat ($K=0$) FRW model, then the scalar 
(Sachs\hs Wolfe) contribution to large scale CMB anisotropies may 
be interpreted as being due to the motion of matter relative
to the surfaces of constant temperature which define the surface 
of last scattering on scales where the instantaneous decoupling 
approximation applies.
 
%%%%%%%%%%%%%%%%%%%%%%%%%%%%%%%%%%%%%%%%%%%%%%%%%
\section*{Acknowledgments}
%%%%%%%%%%%%%%%%%%%%%%%%%%%%%%%%%%%%%%%%%%%%%%%%%
I thank George Ellis, Tim Gebbie, Roy Maartens, Marco Bruni, 
William Stoeger, Alan Coley, Dick Bond, Bernard Carr and the 
referees for useful discussions. This work was supported by the 
FRD (South Africa) and a Dalhousie Postdoctoral Fellowship.
%%%%%%%%%%%%%%%%%%%%%%%%%%%%%%%%%%%%%%%%%%%%%%%%%%
\appendix
%%%%%%%%%%%%%%%%%%%%%%%%%%%%%%%%%%%%%%%%%%%%%%%%%%
\section{Covariantly defined harmonics}
%%%%%%%%%%%%%%%%%%%%%%%%%%%%%%%%%%%%%%%%%%%%%%%%%%
In the standard approach to cosmological perturbations \cite{bi:bardeen,bi:ks}
a harmonic decomposition of the perturbation variables is usually carried 
out using harmonics which are eigenfunctions of the Laplace\hs Beltrami 
operator on the three\hs hypersurfaces of constant curvature i.e. on the 
homogeneous spatial sections of FRW universes. In the covariant approach, 
the fluid four\hs velocity $u^a$ is emphasized rather than an arbitrarily 
chosen spatial slicing, and quantities are defined by projecting 
orthogonal to $u^a$ using the projection tensor $h_{ab}$. 
Covariant harmonics are therefore defined through operators 
constructed with the spatial (orthogonal 
to $u^a$) derivative $\n_a$ which are covariantly constant along the fluid 
flow lines (i.e. independent of  proper time). In this section we will focus 
on scalar harmonics $Q$ which are the eigenfunctions of the covariantly 
defined Laplace\hs Beltrami operator \cite{bi:hawking}:
\be
\n^2Q=-\frac{k^2}{a^2}Q\;,
\ee
where $k$ is a comoving (i.e constant) eigenvalue. If $\nu$ is a
non\hs negative real wavenumber, then for a flat background ($K=0$), it
is associated with the physical wavelengths $\lambda=2\pi a/\nu$, since 
in this case $k=\nu$, however for open models ($K=-1$) the 
spectrum of eigenvalues is given by $k^2=\nu^2+1$ \cite{bi:harrison}.

The scalar harmonic $Q$ can be used to define a vector
\be
Q_a=-\frac{a}{k}\n_aQ
\label{eq:harm1}
\ee
and a trace\hs free symmetric tensor 
\be
Q_{ab}=\frac{a^2}{k^2}\n_b\n_aQ+\case{1}/{3}h_{ab}Q\;.
\label{eq:harm2}
\ee
These harmonics are defined in order to have 
\be
\dot{Q}=\dot{Q}_a=\dot{Q}_{ab}=0,
\label{eq:harm3}
\ee
so that they are covariantly constant along $u^a$. Finally, by taking the 
spatial divergence of (\ref{eq:harm2}) the following relation between 
$Q_{ab}$ and $Q_a$ is obtained \cite{bi:BDE}
\be
a\n^bQ_{ab}=-\case{2}/{3}k^{-1}\left(3K-k^2\right)Q_a\;.
\label{eq:harm4}
\ee
This is needed in section \ref{sec:Newt}.
%%%%%%%%%%%%%%%%%%%%%%%%%%%%%%%%%%%%%%%%%%%%%%%%%%

\end{document}